\documentclass[preprint,amsmath,nofootinbib,tightenlines]{revtex4}
\usepackage{setspace}
\usepackage{graphicx}
\usepackage{bm}
\usepackage{epsfig}
\usepackage{color}
\usepackage{colordvi}

\newcommand{\beq}{\begin{equation}}
\newcommand{\eeq}{\end{equation}}
\newcommand{\bea}{\begin{eqnarray}}
\newcommand{\eea}{\end{eqnarray}}

\newcommand{\mn}{{\mu \nu}}

\begin{document}

\title {Relaxing the cosmological constant in the extreme ultra-infrared}

\author{Rafael A. Porto and  A. Zee}
\affiliation{Kavli Institute for Theoretical Physics, University of California, Santa Barbara, CA 93106}
\affiliation{Department of Physics, University of California, Santa Barbara, CA 93106}

\begin{abstract}
We speculate on the dynamical critical behavior of gravity in the extreme ultra-infrared (EuIR) sector and a mechanism to relax the cosmological constant. We show that in the EuIR the cosmological constant term could be made irrelevant for values of the dynamical critical exponent  $z_{\rm EuIR}$ greater than one. We discuss a possible realization of this idea that connects the relaxation of the cosmological constant to the ratio between the EuIR and IR scales, where the latter serves as the `UV' cutoff of our (ultra long distance) effective theory, with  $z_{\rm IR} \sim 1$. For distances smaller than the IR regime Lorentz invariance emerges. We entertain the possibility that the effective description of the universe may not be Lorentz invariant at much larger scales. We discuss why local physics cannot detect the `natural' value for the density of dark energy below the IR scale, and briefly comment on possible connections with holography.
\end{abstract}

\maketitle

\section{Introduction and Background}

Certainly the Cosmological Constant (CC) problem\footnote{It is usually attributed to W. Pauli the remark that the radius of the world `nicht einmal bis zum Mond reichen wuerde' [would not even reach to the moon] \cite{pauli}.} \cite{zeldo}\cite{swrev} is the most vexing in theoretical physics today. The discovery of dark energy \cite{riess}\cite{perl} has exacerbated the situation into a real crisis in our understanding of gravity. Numerous speculative proposals have been put forward regarding this paradox and discussed in various reviews \cite{witten}\cite{reviewL}\cite{polch}\cite{zyang}\cite{zdirac}. For our purpose here, we will equate the dark energy with Einstein's CC.

The spirit of this paper may be summarized by paraphrasing Einstein's comment that `Physics should be as simple as possible, but not any simpler.' The solution of the CC paradox should be as crazy as possible, but not any crazier. We would like to `think out of the box' but also not stray too far away from it.  We speculate that at large distances gravity might be described by quite a different theory than General Relativity (GR). In particular, its scaling behavior might be governed by a (large) dynamical exponent. This proposal perhaps exemplifies what we have in mind: while it is conceivable, it certainly does not appear to fit into our present day framework for the universe. It is counter-intuitive but not outrageous. 

Our paper is, perhaps by necessity, highly speculative and incomplete. We are not able to back up many of our assertions with detailed calculations. Rather, it should be regarded as a plausible suggestion of how the cosmological term might scale to zero at cosmological distances.

The speculation proposed in this paper is inspired by a suggestive analogy \cite{zdark} with proton decay, first made in \cite{zdirac} and also discussed in \cite{znut}, to which we add several novel ingredients as will be explained below. First, we summarize the analogy.
Suppose that in the pre-quark era we were to estimate the rate $\Gamma$ for proton decay into $e^++\pi^0$. We would simply propose for the effective Lagrangian a dimension 4 interaction ${\cal L}_{pe\pi} \sim f e\bar p \pi$, with $f$ some dimensionless coupling constant. Needless to say, the expected rate comes out way too large, unless $f$ is fine tuned to a ridiculously small number defying the {\it naturalness} dogma. At the time the rate was decreed to be zero by muttering the words `baryon number conservation'. Imagine that in an alternative history $\Gamma$ was measured to be a tiny but non-zero quantity. The resulting crisis would be the analog of what we are faced with today regarding the CC: a `naturally' large quantity was decreed to be identically zero and then measured to be tiny but non-zero.

We now know the resolution of this proton decay paradox. With the invention of quarks by Gell-Mann, Zweig, and Greenberg, the interaction ${\cal L}_{pe\pi}$ gets promoted to a dimension 6 term schematically of the form $\frac{1}{M^2} qqql$. Here $M$ denotes the mass scale of the physics responsible for proton decay, arguably around the GUT scale. In this way the apparently fine tuned coupling $f$ has now a naturally small origin, $f \sim {m^2_p \over M^2}$ for $M \gg m_p$.  Remarkably, the lifting from 4 to 6 is enough to tame an enormous discrepancy with observation. In its quest for longevity, the protons are practically begging to be composite. 

The resemblance with the CC problem is startling. Could we in some way manipulate the CC term
\beq
S=\int dt d^{3}x {\sqrt g} (\Lambda + \ldots ),
\eeq
in a similar fashion? By promoting the mass dimension of $\Lambda \sqrt{g}$ we could make the CC term effectively less important at large distances, that is, less relevant, compared to the Einstein-Hilbert term $ \sqrt{g} R$. One difficulty is that these two terms are manifestly made of the same kind of {\it stuff}. The only difference between the two operators we could see at our present level of understanding of gravity is that $ \sqrt{g} R$ involves spacetime derivatives while $\Lambda \sqrt{g}$ does not. One possible clue is that the Einstein-Hilbert term has to do with curvature while the CC term has to do with volume (perhaps a foam-like structure could distinguish between these two terms \cite{foam}.)

In our analogy, we might have estimated the size of $f$ to be of the order of $\alpha g$ by comparing with the pion-nucleon interaction ${\cal L}_{pn\pi} \sim g \bar p n \pi$. The puzzle can then be turned upside down, by asking why the pion-nucleon coupling $g$ is not also small. The resolution relies on the fact  that  ${\cal L}_{pe\pi}$ and ${\cal L}_{pn\pi}$ eventually turn into rather different type of operators, with $\bar p n \pi$ morphing into ${\bar q}q A$ (with $A$ the gluon field), with the scaling dimension remaining at 4.  

We could imagine in our alternative history that the distinction between hadrons and leptons only became clear much later, but the issue goes beyond whether the fields appearing in the two terms ${\cal L}_{pe\pi}$ and ${\cal L}_{pn\pi}$ are made of the same stuff or not. In the early days of quarks, before Quantum Chromodynamics, theorists could easily have been misled by writing ${\cal L}_{pe\pi} \sim   {\bar q}{\bar q}{\bar q} e {\bar q}q$ as a dimension 9 operator and ${\cal L}_{pn\pi} \sim  {\bar q}{\bar q}{\bar q} qqq {\bar q}q$ as a dimension 12 operator. With ${\cal L}_{pn\pi}$ having a higher dimension than ${\cal L}_{pe\pi}$, the puzzle would have deepened.

In our previous discussion \cite{zdark}, we proposed that somehow $\Lambda \sqrt{g}$ is promoted to a dimension $p$ operator and showed that the CC problem could be solved if $p>4$. Thus, if the proton decay analogy is to be relevant to the CC problem at all, the key issue appears to be the scaling, or in other words, the renormalization group flow, of the various operators describing gravity.\\ 

To carry the discussion further, we have to add some new ingredients to this scaling proposal, as we will discuss in this paper. One new ingredient is the dynamical critical exponent $z$. Being non-relativistic, condensed matter systems  scale differently \cite{zexp}  in time and space: near a critical point, the system scales under the transformation 
\beq
\label{scaling}
t \rightarrow b^{z} t, ~~~ x \rightarrow b x .
\eeq
Lorentz invariance would have restricted $z$ to be equal to 1. We suggest below, however, that in the EuIR regime gravity may break Lorentz invariance and scale with $z$ not equal to 1.\\

Notice that with the dynamical exponent $z \neq 1$ we lose part of general coordinate invariance and so we are indeed speculating that Einstein's venerable theory might break down at cosmological scales. Moreover, as we are about to describe in this paper, once we allow ourselves to depart from Lorentz invariance another relevant feature in our proposal is the break down of locality\footnote{As it has been argued in other contexts (such as black hole evaporation), nonlocality may be an expected feature of a quantum theory of spacetime (see for instance \cite{gipo} and references therein).} in the EuIR due to a contribution in the effective action scaling as $|\omega|^{2/z}$.\\

Much of the literature on quantum gravity has been focused on the UV regime, say from the Planck scale down to the TeV scale. On the other hand, our empirical knowledge of gravity has been restricted to the IR regime, which for our purposes here we will define as ranging from laboratory scale to galactic scale. Assuming that the standard discussion of dark matter is correct, we know that (Post-)Newtonian gravity has been verified up to galactic scales ($\sim 1$ to $\sim 10^{3}$ kiloparsecs), and possibly even up to the scale of galactic clusters. In truth, we know very little about physics in the EuIR regime, which we define as ranging from galactic scales to the scale of the visible universe ($\sim 10^4 {\rm Mpc}$), namely cosmological distances, in which regime one is more or less free to speculate about significant deviation from the standard picture as was emphasized in some discussions of the dark energy\footnote{For example, in some speculations (perhaps even more far-out than what we present in our paper), various authors \cite{nima} suggested the breakdown of causality or the emergence of a five dimensional universe \cite{5dlambda} in the EuIR.}.\\

It is important to distinguish our proposal from Horava's interesting proposal \cite{hor1, hor2}  that gravity may exhibit dynamical scaling with $z=3$ in the UV, flowing to the relativistic value $z=1$ in the IR. Horava's theory has attracted considerable attention \cite{horavath} since it has the remarkable property of being power counting renormalizable. Although the particular realization in \cite{hor1,hor2} may not recover Einstein's gravity in the IR \cite{critics}, the idea of a non-trivial dynamical scaling in the high energy regime remains an open possibility.

We emphasize that we differ from Horava in that he considers the flow from the UV to the IR, while we are interested in the flow from  the IR to the EuIR. In this note, we will not speculate on the behavior of gravity in the UV, which as we just said has been the focus of the literature on quantum gravity for decades \cite{qgreviews}, whereas the EuIR regime has only started to receive some attention.\\

In principle it would be possible to follow a construction similar to that in \cite{hor1,hor2}. However, to flesh out the new ingredients in our proposal, we will restrict ourselves in this note to a toy model where we deal basically with the dynamics of the scale factor. In our scenario of the dynamical scaling flow, $z$ flows from some value $z_{\rm UV}$ (perhaps different from 1 or perhaps not) in the UV to $z_{\rm IR}=1$ in the IR, but then it flows back to some value $z_{\rm EuIR} \neq 1$ at cosmological distances. Here we will focus on the possibility that $z_{\rm EuIR} \gg 1$. 

As in other scenario, Lorentz invariance emerges due to relevant deformations that drive $z$ to 1 in the IR. In this paper, however, we approach this value from the ultra large distances. As we shall see, CC term can be rendered irrelevant in the EuIR regime controlled by the physics of a fixed point with a dynamical exponent different from one. \\

We will explore two realization of this idea and leave for the future the construction of more realistic models. First, we will follow Polyakov \cite{poly} and study metrics of the type (in Euclidean space) $g_{\mu\nu} = \phi^2 \delta_{\mu\nu}$, with the new ingredient of a non-trivial dynamical scaling. We shall show that the CC term  turns off in the EuIR faster than logarithmically (as it would be the case for $z=1$, implicit in \cite{poly}). As in \cite{poly}, the vacuum of the theory lies at $\phi=0$, and therefore in this example our universe (slowly) rolls down towards this state. To approximate a more realistic model, in our second approach we will consider a non-trivial potential for the conformal factor (suggested by the well known existence of negative norm states\footnote{Technically speaking this is due to the non-compact nature of the Lorentz group, which we are no longer assuming as  a symmetry of the theory.} around $\phi=0$) to induce a non-vanishing expectation value, such that spacetime is spontaneously generated. For simplicity we will consider de Sitter (dS) space. Similarly as in the flat case, a CC term induces interactions that become irrelevant in the EuIR. Surprisingly, in both cases we can show that the (potentially harmful) induced mass term does not suffer from a hierarchy problem, basically due to the lack of Lorentz invariance as we run into the EuIR, which effectively decouples the large corrections from the IR.\\

An important new ingredient is the intrinsic role played by dS spacetime in our second scenario. In physics we often implicitly assume continuity. A striking exception \cite{Wqft} is the physics of gauge bosons where, as the massless limit is approached, some extra degrees of freedom may become strongly coupled and remain in the spectrum. In particular, we know for a fact that the universe on cosmological scales is dS, not Minkowski. As we go from Minkowski spacetime to dS the symmetry group expands from the Lorentz group $SO(3,1)$ (more accurately, the Poincar\'{e} group $SO(3,1) \times T(4)$) to the dS group $SO(4,1)$. Physics is different no matter how small the CC. We entertain the nagging suspicion that this discontinuity may have something to do with the resolution of the CC problem.\\ 

An obvious question that arises in our construction is the following: How does local physics manage to hide the large values for the density of dark energy in the IR? We will discuss the solution to this puzzle in somewhat more detail later on. The main observation is that the extension for the (local) free falling frames is determined by the universe as a whole instead of the local patch, in the spirit of a Machian approach. We will then discuss briefly the inclusion of matter into this new picture, and plausible connections with holography.

\section{Dynamical critical phenomena in the Extreme ultra-InfraRed}

It is generally assumed that the fundamental laws of nature are relativistic invariant, so that $z=1$ from Planck to cosmological distances.  On the other hand, it is also possible that Lorentz invariance  only emerges as a residual (accidental) symmetry at intermediate scales. In this note we propose that Lorentz invariance and $z=1$ may be a transient phenomenon, peculiar to the scales that have been experimentally explored. More specifically, motivated by an emergent description of spacetime, we will consider $z \neq 1$, and a non-standard (nonlocal in time) kinetic term in the effective action for the scale factor of the universe such that the (non-derivatively coupled) volume factor becomes irrelevant in the EuIR.

Transient and emergent symmetries are common tokens in condensed matter systems, where the symmetries of the underlying theory are often broken and where novel symmetries may emerge in the effective description at larger scales. Our philosophy is to construct a similar effective description of the universe, hoping that ultimately a fully quantum description of spacetime (perhaps along the lines of \cite{DaSean}) will reproduce the critical behavior in the EuIR that we are about to discuss, in particular its nonlocal features.

Although our goal is to relax the value of the CC (in the same way Horava's aim is to remove the UV divergences in quantum gravity), the notion of transient and emergent symmetries in gravity is an interesting venue per se. Similar ideas have appeared in the literature (see for instance \cite{Wen,Volovik}), however, the novelty of our approach lies on the non-trivial dynamical critical behavior that may occur at ultra large distance scales.

\subsection{Screening the Cosmological Constant}

In an attempt to  go {\it `beyond space and time'} Polyakov \cite{poly} has suggested a mechanism to screen the CC at large distances (see also \cite{jack}.) The main idea in his proposal is to consider a conformal flat universe (in Euclidean signature)
\beq
g_{\mu\nu} = \phi ^2 \delta_{\mu\nu},
\eeq
for which the Einstein-Hilbert action takes the form 
\beq
\label{pol}
S_{\rm E}= \int d^4x \left(\frac{1}{2}M_{\rm Pl}^2(\partial \phi)^2 + \frac{ \Lambda}{M_{\rm Pl}^2} \phi^4\right).
\eeq

In (\ref{pol}) the unwanted minus sign in front of the kinetic energy term has already been removed via analytic continuation $\phi \to i\phi$ according to the Gibbons-Hawking-Perry prescription \cite{GH}. 
Here and henceforth we choose units with $G=M_{\rm Pl}^{-2}=1$.  In these units $\Lambda \sim 1$ and positive (we will occasionally restore factors of $M_{\rm Pl}^2$ for clarity.) In the above expression $\hbar, c$ are set to one, and $x^0=ct$ from Lorentz invariance. In this note, since we will depart from Lorentz invariance, in general we will use coordinates $(t,x^i)$. 
Notice that now the vacuum lies at $\langle\phi\rangle=0$. In our next example we will argue that the minus sign instability suggests a non-zero vacuum expectation value for $\phi$. 

Within this setting Polyakov argues that the IR quantum fluctuations become relevant and the CC is logarithmically screened to zero just as in a standard $\phi^4$ theory. In fact, given that a logarithmic running would not lead to sufficient screening, it has been suggested that the theory is trivial in the sense of Wilson \cite{jack}, which would also imply a (non-perturbatively) vanishing CC.  Indeed, we see that this scheme provides a concrete realization of the proposal in \cite{zdark} in the conformally flat subspace of possible geometries, but just barely since the CC term has been promoted to have mass dimension $p=4$ while $p>4$ was desired.

One argument against Polyakov's scheme is that with general covariance the
conformal fluctuations $\phi$ may represent purely gauge artifacts. After all, we know that the graviton does not contain a physical spin-0 degree of freedom. Polyakov argues that the screening effect is however non-peturbative, since it is caused by fluctuations of the metric around zero. This is also very appealing, since it bears in a primitive notion of emergent spacetime from a non-metric state with $\phi=0$. 

This objection will not apply to us however, since, as we will discuss in detail below, we propose to abandon general covariance in the EuIR. We suppose that in the EuIR Einstein's theory is replaced by another theory in which the dynamical exponent $z$ is not equal to 1 and so $\phi$ could well be physical. Although we will keep this basic picture, our vacuum will ultimately be moved away from the non-metric state\footnote{Notice that already in Polyakov's construction, we could invoke the Coleman-Weinberg mechanism to move $\phi$ away from zero. This is extremely appealing. However again it requires non-perturbative effects, since as it is well known, the minimum of the Coleman-Weinberg potential lies far away from the validity of the one loop approximation. Another point, to which we will come back later, is the fact that a mass term can also be generated from quantum effects.}.	

\subsection{Making the Cosmological Constant Irrelevant}

As we mentioned above, the CC term turns into a marginal interaction, which could be screened away logarithmically. One natural question then arises: Can we turn the CC into an irrelevant interaction? Surprisingly the answer is yes, but we need to `go beyond space and time' differently from Polyakov. In a nutshell, in contrast to Polyakov, we go beyond spacetime by splitting spacetime into space and time.\\

Thus, we now differ from Polyakov by abandoning Lorentz invariance.  As we mentioned before, in condensed matter systems one considers the behavior of the system under the scaling (\ref{scaling}). Since we work with a conformally flat spacetime\footnote{Notice that, even though the `background' spacetime is Lorentz invariant, the solution for $\phi(u,x)$ will break the symmetry already for an `empty' universe with a CC.}
\beq
\phi^2(u,x)(du^2+d {\vec x}^2),
\label{conftime}
\eeq 
it is natural to interpret (\ref{scaling}) as
$
u \rightarrow b^{z} u, ~~ x \rightarrow b x ,
$
with $u$ the `conformal' time. The scaling (\ref{scaling}) $\omega \rightarrow b^{-z}\omega, ~~k\rightarrow b^{-1}k$ forces us to replace the first term in (\ref{pol}) by a free action of the form\footnote{Note that in principle we also have a choice between $i\omega$ (Schr\"odinger) and $|w|$ at $z=2$.} (again in Euclidean signature)
\beq
\label{cmz}
S_{0}=\int \frac{d^3kd\omega}{(2\pi)^4}\left(k^2 + |\omega|^{2/z}\right) |\phi(\omega,k)|^2,
\eeq

Let us restore momentarily high-school dimensions in (\ref{cmz}) and write
\beq
S_{0}=\int \frac{d^3kd\omega}{(2\pi)^4}\left(k^2 +\frac{\omega_*^2}{c_{*}^2} |\frac{\omega}{\omega_*}|^{2/z}\right) |\phi(\omega,k)|^2,
\label{cmz2}\eeq
where we have introduced $\omega_*$, the frequency scale at which the non-trivial dynamics sets in and $c_{*}$, a `coupling constant' to convert frequencies into inverse distances; later in section E we will introduce the constant that will play the role of the `speed of light'. We identify $\omega_*/c_{*} \sim 1/l_{\rm IR}$, where $l_{\rm IR}$ correspond to the galactic scale. As mentioned earlier, we take it to be 

\beq \label{lir} l_{\rm IR} \sim 1-10^3 ~{\rm kiloparsecs}. \eeq

In what follows we will set $\omega_*$ and $c_{*}$ to 1 unless otherwise noted.

We will work out the Fourier transform of the `temporal' piece in $S_{0}$ shortly. For now, let us focus on the Fourier transform of the `spatial' piece: $\int d^{3}x du (\partial_{x} \phi)^{2}$. Requiring this to be invariant  fixes the scaling \beq  \phi(u,x) \rightarrow b^{-(1+z)/2} \phi(u,x),~~ {\rm or} ~~ \phi(\omega,k) \rightarrow b^{(5+z)/2} \phi(\omega,k).\label{scalphi}\eeq  

It follows immediately that the term $S_{\lambda}=\int {d^3xdu} ~\lambda \phi^4(u,x)$ representing the CC scales like
\beq
S_{\lambda} \to  b^{3+z-2(1+z)}S_{\lambda}=  {\frac {1}{b^{z-1}}} S_{\lambda}
\eeq
in four dimensions, and therefore it becomes irrelevant at long distances for $z>1$. 

By splitting spacetime into space and time, we have managed to make the CC scale as 
\beq
\Lambda(l) \sim \left(\frac{1}{l}\right)^{z-1}
\label{result}\eeq
at large distances $l \rightarrow \infty$ (for a Lorentz invariant theory $z=1$ and we recover Polyakov's observation that the $\lambda \phi^4$ coupling is marginal.) The larger the dynamical critical exponent, the more irrelevant $\lambda$ becomes, thus suggesting a natural mechanism to screen the CC in the large $z$ limit. The crucial point is that, after integrating out the {\it short distance} physics, the interaction becomes irrelevant in the EuIR.

Moreover, since there is no symmetry that protects $z$, quantum corrections would modify the scaling laws, producing non-integer $z$'s in general. In this scenario it is not necessary to resort to loop effects to break scale invariance, since we are already modifying our `unperturbed' theory to include a non-Lorentz-invariant dynamical critical behavior.\\ 

Notice that (\ref{pol}) has dS space as solution, that is $\phi(u)=\frac{1}{Hu}$, with $H \sim \sqrt{\Lambda}$ the Hubble constant\footnote{Recall that with Euclidean signature dS has the topology of a sphere, which may no longer be the case for a theory with a non-trivial dynamical exponent since dS is no longer a solution.}. 
With the  $\Lambda \phi^4$ term marginal for $z=1$, as in (\ref{pol}), $H$ will turn out to be absurdly large. In our scenario, however, the $\Lambda \phi^4$ term will be driven to zero in the EuIR, not logarithmically as in \cite{poly}, but by the scaling law (\ref{result})
\beq
\Lambda({\rm EuIR}) \sim \Lambda({\rm IR}) \left(\frac{l_{{\rm IR}}}{l_{{\rm EuIR}}}\right)^{z-1},
\label{scalinglaw}
\eeq
where \beq \label{leuir} l_{\rm EuIR} \sim 10^4 {\rm Mpc}\eeq represents the scale of the universe. Thus, combining (\ref{lir}) and (\ref{leuir}) we have $l_{{\rm EuIR}}/l_{{\rm IR}}\sim 10^4-10^7$. To screen the CC to the desired value we would need a factor of $\sim 10^{-120}$, which we can achieve with reasonable values of $z$, namely \beq z_{\rm EuIR} \sim 20-30. \eeq  

This number can be relaxed if instead of the Planck scale, we take $\Lambda({\rm IR})$ to be set by the TeV scale.
Therefore, we have found a mechanism by which spacetime remains essentially flat at large scales, hence potentially solving the cosmological hierarchy problem.

\subsection{Non-locality in time and slow roll} 

The action $S_{0}(x, u)$ is local in space but not in time since in general the kernel
\beq
\label{ker}
K(u_1-u_2) \equiv \int d\omega |\omega|^{2/z} e^{i\omega(u_1-u_2)}
\eeq
is not given by $\delta(u_{1}-u_{2})$ differentiated an appropriate number of times. Instead, we have a `kinetic' term of the form\footnote{Except for the Lorentz $z=1$ and Schr\"odinger  $z=2$ cases, where we recover a local kinetic term.}
\beq
\label{kinet}
\int du_1du_2 \frac{\phi(u_1)\phi(u_2)}{|u_1-u_2|^{1+2/z}},
\eeq
Therefore it is clear that the solution with a non-trivial dynamical exponent, namely $\phi(u)$ for the kernel in (\ref{ker}), will not be equivalent to the Lorentz invariant case. In particular, it will not 
be of the form $\phi(u) \sim 1/\sqrt{\Lambda({\rm EuIR})}u$, thus we do not wind up in a dS universe with a small CC. What we have shown instead, is that the physics of a fixed point with $z_{\rm EuIR} \gg 1$ renders the `volume' factor $\Lambda\sqrt{g}$ irrelevant. 

In the EuIR we expect the function $\phi(u,|\vec{x}|\sim l_{\rm EuIR})$ to remain essentially constant in $u$, rolling down (very slowly) towards the $\phi=0$ vacuum in an essentially flat $\phi^4$ potential. In addition, our kinetic term is  non-local in time, of the type (\ref{kinet}). As discussed in \cite{nima}, in principle it is possible that the breakdown of causality in the EuIR is a necessary ingredient to solve the CC puzzle. Here we are adding a new ingredient to the mix, namely nonlocality in time. However, we may still restore causality by considering by fiat only the retarded propagator in (\ref{ker}). Solving the integral equation 
\beq
\int^u du' \frac{1}{|u-u'|^{1+\frac{2}{z}}} \phi(u') \sim \Lambda({\rm EuIR}) \phi(u)^{3}
\eeq
we obtain a `crawling' behavior of the sort (for large $z$)
\beq 
\phi(u, |\vec{x}| \sim l_{\rm EuIR}) \sim \frac{1}{\sqrt{\Lambda({\rm EuIR})} u^{1/z}}.
\eeq

Notice that conformal time $u$ and co-moving time $t$, defined by
\beq
dt^{2}+a(t)^{2} d{\vec x}^2
\eeq are basically the same for large $z$, since
\beq
t \sim \int \frac{du}{u^{    \frac{1}{z}     }          }       \sim u^{1-{\frac{1}{z}}     }
\eeq
Therefore our universe will remain essentially flat
\beq
a(t)=\phi(u) \sim \frac {1}{   u ^{    \frac{1}{z}     }   } \sim \frac{1}  {  t^{   \frac{1}{z-1}     }         }
\eeq although it shrinks (albeit very slowly for large $z$) as it drives towards the bottom of the potential. Notice that in our case, conformal and co-moving time point in the same `direction', whereas in the relativistic (dS) scenario future-infinity lies at $u \to 0$ (where the volume factor, i.e. the CC term, becomes crucial.) In both cases, the `forward' evolution in conformal time is towards the bottom of the potential\footnote{One possibility to remedy this behavior is to consider a negative CC, i.e. $\Lambda<0$, which would produce an unstable potential, and henceforth a growing scale factor.}.  In the next section we will explore a modified potential with a non-vanishing vacuum expectation value for $\phi$. \\

While we have found the CC can be relaxed at large scales, at this stage ours is a  simplified, yet very suggestive model. What we have shown is that the dynamical critical behavior of the universe in the EuIR could relax away the large (local) values of the CC.  However, our effective description is only valid towards the late stages of the evolution (possibly after matter dominance) when $\phi \ll 1$, and $\Lambda(\phi) \ll 1$. In some sense, ours may be described as a `technically natural' solution to the CC problem, since we rely on the existence of a deeper theory where a sufficiently flat background emerges dynamically (i.e. solves the full equations of motion), and then remains essentially stable as shown by the effective action description, at least for the perturbations of the scalar factor $\phi$ (that we have argued are not gauge artifacts once we depart from Einstein's theory).\\

Another possibility is to think of the background as given by some underlying mechanism (perhaps even as a fundamental {\it stage} of nature) and gravitational forces as an emerging dynamical field living on the (coarse-grained) cosmological background, such that its dynamics resembles standard GR in the IR (possibly with $z \sim 1$), at least at the Post-Newtonian level around flat space. Rather than the latter, an appealing scenario is to take dS as a background, where $\Lambda(\rm EuIR)$ may be a spontaneously generated scale (which we may also allow to vary with time, in Planck units.) We analyze  this other option in what follows.

\subsection{Spontaneously generated de Sitter spacetime} 

In our previous model the universe `crawls' towards the vacuum, namely the non-metric state $\phi=0$. In this section we will consider a modification of this picture where the vacuum of the theory sits at a non-zero value. Thus, one natural modification to Polyakov's idea is to consider dS instead of $\delta_\mn$, and to consider metrics of the form $\phi^2 g^{\rm dS}_{\mn}$, where 
the dS spacetime takes the form (in Euclidean signature)
\beq
\label{desitter}
g^{\rm dS}_{00}dt^2 + g^{\rm dS}_{ii} (\delta_{ij} dx^idx^j)  = dt^2 + a(t)^2 d{\vec x}^2
\eeq
with $a(t) =\frac{1}{H_{\rm EuIR}} \cos(H_{\rm EuIR} t)$ an oscillating scale factor of the universe. In what follows we will assume that $H_{\rm EuIR} \sim \sqrt{\Lambda({\rm EuIR})}$ is determined by the EuIR scale, thus we can consider the period in Euclidean time to be effectively very large. Notice that this background breaks Lorentz invariance (and diffeomorphism invariance) explicitly, which selects the coordinate system $(t,x^i)$ that plays a crucial role in our analysis. Even though we make no further assumptions about the emergence of the dS background, we also entertain the possibility that  $\Lambda(\rm EuIR)$ may be a fundamental constant of Nature (perhaps varying with time), playing a role similar to that played by 
the speed of light.\footnote{Although this is not the road we pave in this paper, we remark that in principle we could also think of a dS extension of GR (i.e. locally dS free falling frames) such that $l_{dS}$, the dS curvature scale (e.g. $[P,P] \sim \frac{1}{l_{dS}^2}$), is promoted to the same status as $c$, the speed of light. Likewise, since $c$ doesn't get renormalized, this theory would provide a (technically) natural solution to the CC problem. In such theory (with a dS invariant coupling with matter) rather than an infinite renormalization of $l_{dS}$, we would get an overall wave function renormalization of $G_N$, similarly as $\alpha_{\rm em}$ in QED (although in `natural' units $l_{\rm Pl}$ would turn out to be absurdly tiny). Related ideas appeared in \cite{dS}, where it has been baptized `de Sitter Relativity'.}\\

The first obvious obstruction to working with a scalar field (our conformal factor) in a dS background is the issue of potentially dangerous IR instabilities \cite{poly}. There are already suggestions that IR fluctuations in GR could become relevant and may drive the CC to zero (see for instance \cite{tsamis}). However, we are not assuming that GR holds at the EuIR scale, therefore as a first approximation (and also ignoring the `Hubble friction'), we will assume that the previous picture does not get modified considerably, except for the inclusion of a mass term, which in principle could also appear from quantum effects (see below.) If we had Einstein gravity we would have a (conformally coupled) mass term of the form $\frac{1}{12} R(g^{\rm dS}) \phi^2 \sim \Lambda({\rm EuIR}) \phi^2$ in the Lagrangian. We will assume a similar contribution in our model. However (motivated by the  ghost-like instability in the GR case), we will take the sign to be the opposite to the kinetic term, thus introducing a instability in the potential. Therefore, we take the latter to be 
\beq
\label{potent}
V(\phi) = \Lambda(\rm EuIR) (\phi^2 - 2)\phi^2.
\eeq

This potential has minimums at $\phi^{\pm}_0 = \pm 1$, and henceforth our dS spacetime is  spontaneously generated. Following the philosophy of the  past section, we can show that the fluctuations of the geometry are effectively small, which provides another (yet more realistic) technically natural solution to the CC puzzle. The steps will mimic our previous discussion, with the exception that the potential in (\ref{potent}) has an instability around $\phi=0$ that will move the minimum towards $\phi=1$, thus producing an emerging (time-dependent) dS background.\\

This is also a very appealing scenario, though somewhat different from what we discussed previously.  In this  `superfluid' phase we expand around the dS vacuum, that is we study (small) perturbations of the form $\phi=\phi^+_0+\varphi$. By assuming a dynamical critical scaling as in (\ref{scaling}) in the $(t,x^i)$ coordinates, and $\varphi$ as in (\ref{scalphi}), it is easy to show that the CC turns into an irrelevant interaction in the EuIR as in the previous case (although now we work with  free-falling (that is, co-moving) time rather than conformal time\footnote{Since we expect Lorentz invariance to emerge in the IR, local observables will follow geodesic motion and measure proper time as in relativistic theories.}.) Furthermore, as we shall show in what follows, the tiny mass term $m_{\varphi}^2 \sim \Lambda(\rm EuIR) \sim (10^{-3} {\rm eV})^4/M_{\rm Pl}^2$, is also natural due to the irrelevance of the $\varphi^4$ interactions. We can now allow ourselves to paraphrase the Condensed Matter jargon, since the picture that emerges here may be also thought of that of a filled  Fermi-sea that we can naively associate to the spacetime background, and `quasiparticles', e.g. $\varphi$, with a (canonical) kinetic term of the form $k^2 + |\omega|^{2/z}$ at the fixed point, and an irrelevant $\varphi^4$ interaction. Notice this resembles the effective field theory (Lorentz variant) treatment of Fermi liquids by Polchinski \cite{joe} (although in that case the holy grail was to study the RG flow in the marginal case and the onset of instabilities (Cooper pairs) in the Fermi surface.)\\

Let us pause for a second and add a few comments here. Notice that in principle we could be more generic and consider metrics of the type \beq g_\mn dx^\mu dx^\nu=\phi^2dt^2+ \chi^2 d{\vec x}^2, \eeq such that our vacuum will take the shape $\langle\chi\rangle \sim a(t)$, $\langle\phi\rangle \sim 1$. The condition $\langle\chi\rangle \sim a(t)$, somewhat resemblance the ghost-condensate scenario of \cite{gcond} (recall that in that case the ghost field picks a time dependent expectation value and the resulting scalar Goldstone mode mixes with its gravitational counterpart.) However, as a simplifying assumption we considered the case of a single scalar perturbation around dS (for example, in the case of GR (and perturbations around flat space) the equations of motion set $\phi=\chi$.) Moreover, in GR these fields are not dynamical (that is the origin of Polyakov's objection to work with the conformal factor, and the arguments about non-perturbative effects around $\phi=0$.) In our case, since we depart from GR, we do not expect the effects discussed above to be just pure gauge artifacts. This on the other hand, brings in the issue of potentially harmful extra degrees of freedom in our theory, as is the case in Horava's gravity, namely an extra (scalar) degree of freedom. In fact, it is even conceivable that our considerations are a particular case of a  ghost-like condensate, with non-trivial dynamical scaling. 

As we indicated before, our kinetic term is nonlocal in time, which may suggest some of these degrees of freedom have been `integrated-out'. On the other hand, it is also possible that our description derives from a deeper (intrinsically nonlocal) structure. Although an essential element in our model, we will not attempt a general treatment in this paper. 

The main point we want to stress is the possibility of a non-trivial (time-dependent) dynamical critical behavior in the universe, and the possibility of relaxing away the large values of the CC.

\subsection{Lorentz Invariance and the hierarchy problem} 
 
Usually it is Lorentz invariance that protects the dynamical exponent $z$ from deviating from 1, but here in our scenario it works the other way around. Our scheme of thought follows rather naturally from the standard procedure of adding a series of relevant operators to the effective action, so that the kinetic energy factor in (\ref{cmz}) actually has the form $\left(k^2 +|\omega|^{2/z} +\ldots \right)$, with the dots represent the effect of the relevant operators, less dominant at small $k$ and $\omega$ than the terms $k^2 +|\omega|^{2/z}$ actually displayed (for example, the term $\omega^{2}$ that would dominate $|\omega|^{2/z}$ (we are as before taking $z>1$) at larger values of $\omega$  and restore Lorentz invariance.) We suppose that around $l_{\rm IR}$ we have the crossover form $\left(k^2 +|\omega|^{2/z_{\rm EuIR}}+\ldots+\frac{\omega^2}{c^2} +\ldots \right)$. From EuIR distances our theory flows into $z=1$ at `short' IR distances, where `short' in this paper means small compared to galactic sizes. 

Here, as promised earlier, we have introduced the parameter $c$ that measures the relative importance of the $\omega^{2}$ term, which evidently plays the role of the speed of light in the $z=1$ regime with the form $\left(k^2 +\frac{\omega^2}{c^2} +\ldots \right)$ . Recall earlier we used dimensional analysis (see (\ref{cmz2})) and introduced the parameter $c_{*}$. A priori, $c$ and $c_{*}$ could be quite different, although the (possibly discredited) naturalness dogma would suggest that they are similar in magnitude, or perhaps even identical. Let us introduce the length scale $l_* \equiv c_*/\omega_*$  and re-write (\ref{cmz2}) as $\left(k^2 +{\frac{1}{l_*^2}}|l_*\omega/c_{*}|^{2/z_{\rm EuIR}}+\ldots+\frac{\omega^2}{c^2} +\ldots \right)$. Evidently, the standard $\frac{\omega^2}{c^2}$ term takes over when $|w| > c/l_*$, and thus the coefficient $l_*$ sets the scale of $l_{\rm IR}$. In general, the speed of light $c$, as well as the coefficients in the other relevant deformations, may have non-zero scaling dimension, as in Horava's theory\footnote{Notice that in principle, if we include more than one degree of freedom, the emergence of Lorentz invariance requires the speed of light for each `species' to be the same, otherwise it wouldn't make sense to define $x^0=ct$ as a kinematical condition.}.\\

Notice that in principle we could also generate a mass term $m_{\phi}^2\phi^2$ via quantum fluctuations so that our kinetic factor becomes $\left(m_{\phi}^2 +|\omega|^{2/z_{\rm EuIR}}+\ldots +\omega^2 +k^2+\ldots \right)$ (where we have rearranged our terms for convenience and set $c$ and $c_{*}$ both to 1, assuming that they are similar or equal in magnitude). Thus, at really long time, the mass term would dominate the term we want $|\omega|^{2/z_{\rm EuIR}}$. Moreover, in principle the mass term would be a  relevant operator which gets renormalized. It could also drive the minimum of our potential in (\ref{potent}) away from $\langle \phi \rangle =1$. Therefore, naively we would expect a hierarchy problem to set in, thus driving $m^2_\phi(\rm EuIR)$ to the `UV' scale of our EuIR effective theory, namely $\Lambda(\rm IR)\sim M_{\rm Pl}^2$, which would be disastrous. 

Here is where the critical behavior turns out to be a savior, and unravels a very interesting aspect of the breaking of Lorentz invariance in our setting, namely the $\phi^4$ interaction is irrelevant in the EuIR. If we set up the `UV' or `short distance' cutoff of our effective EuIR theory at $\sqrt{\Lambda(\rm IR)} \sim M_{\rm Pl}$, quantum fluctuations will drive our mass term in the EuIR as 
\beq
\delta m_{\phi}^2 ({\rm EuIR}) = \lambda({\rm EuIR}) \Lambda({\rm IR}) \sim  \left(\frac{l_{   {\rm IR}  }   }{     l_{    {\rm EuIR}    }    }\right)^{z-1} \Lambda({\rm IR})\sim \Lambda({\rm EuIR}) \sim \frac{(10^{-3} eV)^4}{M_{\rm Pl}^2},
\eeq
giving rise to a very tiny mass indeed. Notice that this follows basically from power counting. In fact, we could compute the correction  with a Lorentz invariant propagator since the dangerous contributions from the loop come from the very short distance (that is, distances of the order  $l_{\rm IR} $) physics where we assume $z=1$. 
Therefore, we do not have a hierarchy problem, as long as the CC interaction becomes irrelevant in the EuIR. Notice that this is a way to control the size of relevant operators once would-be marginal terms become irrelevant. The addition of matter fields may alter this picture, and induce corrections that could scale with the matter density. We touch upon this point next.

\subsection{Friedmann-Lemaitre-Robertson-Walker Universe}

The inclusion of matter in our picture could in principle modify our conclusions, for starters we do not live in a dS universe but in a Friedmann-Lemaitre-Robertson-Walker (FLRW) type. Here we could take a  phenomenological approach and consider \beq a_{\rm FLRW}(t) \sim a_0 t^{\frac{2}{3(w+1)}}\eeq
in (\ref{desitter}). 
The inclusion of matter selects a natural (isotropic) frame, thus also breaking Lorentz Invariance\footnote{This is nicely observed via Doppler shift in the CMB, the so called `dipole effect'. Indeed, the Solar System is moving at  $\sim$ 370 km/sec relative to the FLRW frame.}. We could follow our previous steps and argue for a non-trivial dynamical critical exponent to assure the stability of this solution. Notice for instance that, naively, the local matter density $\rho$ is much smaller than that of $\Lambda(\rm IR)$, hence it seems difficult to imagine it could play a role at the EuIR scale. Unfortunately assuming a FLRW background would leave little predictive room, and moreover we would not have a natural way to associate $w$ with the stress energy tensor as in Einstein gravity (i.e. $p = w \rho$). A more ambitious idea is to retain the essential features of our previous section and introduce matter fields in our Lagrangian via extra (time-varying) degrees of freedom, after all, only the total energy is conserved (i.e. $\rho a^3$). In such a modified picture one could in principle attempt to find solution of the equations of motion that would resemble our FLRW universe, perhaps without destroying the nice features of the model. In essence, we should keep in mind that the only time-independent factor in our Lagrangian is the one associated to the CC, and therefore it is not unthinkable that a non-trivial value of $z$ could turn it into an irrelevant interaction without (significantly) modifying the other features of the model (at least at scales where the universe is extremely large.)  We will study this in more detail in the future.

\section{The global picture}

The most pressing obstacle to claim triumph is the naive expectation of potentially large corrections in the local dynamics obeying Lorentz invariance, and for which $z=1$, due to a Planck-size CC of order $\Lambda(\rm IR) \sim 1$ in the IR. Even if we were to suppose that Polyakov's argument would hold around flat spacetime, the slow logarithmic running would not save us from catastrophic results. Moreover, we would expect at least an energy density of the order of $({\rm TeV})^4$ around matter from the Higgs condensate.  Naively we would expect to introduce corrections to the equations of motion in the solar system via a potential of the form $-\Lambda({\rm IR}) r^2$. 
So, how is it possible that the large energy density of the vacuum does not affect the local physics?\footnote{We thank J. Polchinski for a discussion on this point.}\\ 

In principle one may argue that, even though we have managed to produce an almost flat universe, if anything resembling Einstein's gravity were to survive in the IR (and below), the effective action for gravity would be subject to a very large CC term (if not via the `tadpole' since we have a quasi-flat backround, at least from the corrections in the propagator.) This could potentially introduce large corrections to the geodesic motion of local bodies. However, if true, this would also lead local observers `at rest' in an essentially flat universe to conclude theirs is not an inertial (i.e. `free-falling') frame, violating the equivalence principle. Although this is certainly a possibility, the picture we have in mind points in a different direction.

In Einstein's theory the presence of a CC requires an expansion of Einstein equations around a shifted background for which $R=4\Lambda$.  But if we would consider $R=4\Lambda(\rm IR)$ locally, we will immediately run into a tension with the fact that  `globally' $R \sim \frac{\ddot a}{a} \sim \Lambda(\rm EuIR) \ll \Lambda(\rm IR)$. 
Therefore, it is clear that we require a mechanism by which the effective descriptions of the universe at different scales can be merged `smoothly' into a unified theory of space and time (plausibly this requires some sort of `short-distance/long-distance mixing' in the underlying theory.)\\

In our picture, even though each local region is attempting a bursting expansion with a Hubble constant $H({\rm IR}) \sim 1$, the whole space manages to bypass naive scaling arguments and to evolve according to a much smaller Hubble parameter. In a nutshell, this is the basic idea. Imagine we live in a sphere with curvature $1/l_{dS}^2$. As long as we walk around a neighborhood of size $l \ll l_{dS}$ we will feel as if we are roaming around flat space, up to corrections of order $(l/l_{dS})^2$. In other words, locally we will be free-falling in an approximately inertial frame of size $\sim l$. Imagine now that locally each region of spacetime, of size $l$, is trying to rip apart and grow exponentially fast. If this symphony continues in sync all over spacetime, the whole sphere will react to the will of its subparts.  However, if somehow at larger scales the collective phenomena were to relax away this impetuous rebellion, each local region will somehow be sensitive to the next in line, and `press' against each other to avoid the large explosion\footnote{Amusingly, this is reminiscent of the surface of a virus \cite{virus}. Loosely speaking, we can imagine a sort of  holographic pressure from the `boundary' of each local region.}. Within this scenario local geodesic-deviation will still be proportional to $\ddot a/a$, and therefore totally unobservable for local clocks and rulers, as we would expect from the equivalence principle. 

In a sense we are introducing a Machian type of universe, where the size of the local inertial frame is determined by the Universe as a whole, and not by local physics. Nevertheless, all the physics in the local frames will remain as we know it, in particular Lorentz invariance will be recovered, with Post-Newtonian corrections around (essentially) a flat background. \\

\subsubsection{The holographic scenario}

It is instructive to contrast our picture with the holographic dark energy scenario \cite{Hsu} proposed by Hsu. First of all, we avoid the objections raised towards the holographic dark energy picture. Those objections would apply only if there is a single region of size $l \ll l_{\rm EuIR}$ filled with dark energy. Since we expect the whole universe to form a uniform vacuum, we cannot treat each local piece independently in a holographic setting, and therefore local bounds do not apply. Indeed, as we mentioned earlier, the effective action for the universe of (\ref{cmz}) turns out to be non-local in time. Surprisingly, it remains local in energy, which still allows us to study its RG behavior. 

The main objection to the holographic dark energy picture is the fact that, taking the infrared cutoff at the scale of the universe implies $\Lambda \sim 1/l^2$, with $l$ the size of the horizon. In a matter dominated epoch, $l \sim a^{3/2}$ and hence $\Lambda \sim 1/a^3$, giving a dark energy whose equation of state resembling that of pressureless matter, and therefore ruled out by the current bounds $w < -0.78$ \cite{Hsu}. In other words, the CC looks pretty much constant since matter started to dominate the universe. 

Although we haven't yet studied the inclusion of matter in detail, it is interesting to discuss the variations of our scheme versus the naive application of the holographic idea.\\ 

In our scenario we could move away from the fixed point by including relevant deformations such that the dynamical exponent $z$ itself depends on distance. We may consider a scaling law of the form 
\beq
z(l) \sim \left(\frac{l}{l_{\rm IR}}\right)^{\alpha}.
\eeq
In the above equation $\alpha$ is some unknown positive power for $l>l_{\rm IR}$ so that starting from $z(l_{\rm IR})\sim1$ the dynamical exponent $z$ could attain large values at cosmological scales. Thus, for $z(l_{\rm EuIR})\gg 1$, we replace  $\Lambda(l) \sim l^{-z(l)}$ from (\ref{result}) and (\ref{scalinglaw}) effectively by, 
\beq
\Lambda({\rm EuIR}) \sim  \Lambda({\rm IR})\left(\frac{l_{\rm IR}}{l_{\rm EuIR}}\right)^{z(l_{\rm EuIR})} \sim \Lambda({\rm IR}) e^{-\alpha^{-1}z(l_{\rm EuIR})\log z(l_{\rm EuIR})}.
\label{screen}
\eeq

Here we have two possible roads to take. Obviously, we can always consider the ratio $l_{\rm EuIR} / l_{\rm IR}$ to be constant in time yielding a true CC. Another option is to consider $l_{\rm IR}$ as a fixed and large scale (perhaps the size of the universe around the starting point of the matter dominated era), and apply the above equation only for $l > l_{\rm IR}$. This has an obvious drawback, that is a gigantic CC at the time where the universe's size is $l \sim l_{\rm IR}$. There is an obvious solution to this problem, which also offers us with another implementation of our idea. That is, we could apply the holographic bound to each local (quasi-flat) region of size $l_{\rm IR}$, hence $\Lambda({\rm IR}) < \frac{1}{l_{\rm IR}^2}$. This will significantly reduce the starting point of the scaling law.  In this alternative scenario we may not need such a large values of $z$ to obtain the desired effect. Recaping, we can either assume that the ratio of scales remains fixed, in which case the CC is a true constant (up to some other underlying dynamics), or instead we can use holography once again and consider the following scaling picture 
\beq
\Lambda(l) \sim \frac{1}{l_{\rm IR}^2} \left(\frac{l_{\rm IR}}{l}\right)^{z(l)-1}\sim \frac{\rm const}{l^{p(l)}},
\eeq
where we restored $z$, and its scaling behavior, as an independent variable and ${p}(l)=z(l)-1>0$. Notice that $p \sim 2$ ($z \sim 3$) restores the (global) holographic dark energy ansatz. 

\section{Conclusions}

{\small {\it `...what has the universe got to do with it? You are in Brooklyn! Brooklyn is not expanding!'} \\
Alvy's mother \cite{ahall}.}\\

The nature of dark energy remains a mystery. In science, however, we thrive on  mysteries and confusion. The accelerated expansion of the universe is a well defined fundamental question to be answered. There is always the possibility of a major paradigm shift, which may be required to solve the CC problem and ultimately to understand gravity at disparate scales. Our attempt in this note is to explore the possibility of significant departures from the standard lore in order to gain insights that may lead us towards a satisfactory understanding of the CC puzzle. One possibility is to study departures of Lorentz invariance in the extreme-ultra-infrared\footnote{More recently, a `healthy' extension of Horava's model has been proposed that reduces to a Lorentz-violating scalar-tensor theory at low energies \cite{sibi}.}, and the presence of a non-trivial dynamical critical exponent.\\ 

Violations of Lorentz invariance have been exhaustively studied in the literature \cite{Lorentz invarianceV}. 
It is certainly possible for Planck scale physics to break Lorentz invariance in the low energy theory via higher dimensional operators\footnote{We should note however that recent results from the Fermi collaboration \cite{fermi} indicate that Lorentz invariance may hold all the way to the UV. This result implies that the coefficients of dimension five operators (suppressed by the Planck mass), that modify the dispersion relation and violate Lorentz invariance, are `unnaturally' small. Or in other words, the relevant scale for Lorentz violation is much higher than the Planck mass, whereas no significant constraints appear for dimension six or higher.}. However, in such a scenario one is faced with the task of explaining away Lorentz violating terms of dimension 3 and 4 that could also be generated \cite{myers}. The point is that, unlike baryon number violation say, it is easy to write down low dimensional terms that violate Lorentz invariance\footnote{One possibility is the existence of extra symmetries that protect Lorentz invariance from breaking via lower dimensional terms.}. In any case, we would need to know the details of the UV completion in order to adequately understand how these terms are `naturally' suppressed.\\

In our scenario we may imagine a Lorentz invariant universe up until the EuIR scale where the new (nonlocal) dynamics controlled by a dynamical critical exponent starts to set in.
In our model the main features are the breakdown of (time) locality and general covariance, that allowed us to render the CC, i.e. the volume factor, irrelevant (in principle we could also consider spatial nonlocality by adding terms scaling as $|k|^{2/y}$.) Admittedly, we could not substantiate many elements of our picture, but it may also be appealing to believe that new physics can emerge at large distances. Einstein curved space and time. Here we are suggesting that the logical next step might be to endow space and time with some `substance', such as would be the case in some kind of `foamy' picture of emergent space and time.\\

It would be a bit disappointing if dark energy proves to be merely due to mundane reasons, such as the presence of some scalar fields, and ultimately fine tuned\footnote{A this stage it may also be inappropriate to be overly concerned with the hierarchy problem. In fact, this sort of consideration may remove long-lasting objections and, in turn, have a strong impact in other realms of physics, such as electroweak symmetry breaking, neutrinos and dark matter \cite{zeesilv,ph1,ph2}.}.  
What we presented here is admittedly highly speculative and rather far from a complete theory, but which we hope will provide a starting point for fruitful discussions.
  
\acknowledgments

We would like to thank David Berenstein, Raphael Flauger, Sean Hartnoll, Joao Penedones and Joe Polchinski for stimulating discussions. This work was supported by the Foundational Questions Institute (fqxi.org) under grant RPFI-06-18 and by the NSF under Grant No. 04-56556.

\end{document}